\documentclass[final,3p,times]{elsarticle}

\usepackage{amssymb}

\usepackage{lineno}
\usepackage{hyperref}
\usepackage{calc}
\usepackage{soul}
\usepackage{xcolor}

\makeatletter
\def\ps@pprintTitle{%
 \let\@oddhead\@empty
 \let\@evenhead\@empty
 \def\@oddfoot{\centerline{\thepage}}%
 \let\@evenfoot\@oddfoot}
\makeatother

\begin{document}

\begin{frontmatter}



\title{Unexpected Field Evaporation Sequence in $\gamma$-TiAl}

\author[inst1]{Jiayuwen Qi}

\affiliation[inst1]{organization={Dept.~of Materials Science and Engineering, The Ohio State University},
            city={Columbus},
            state={OH},
            country={USA}}

\author[inst2]{Fei Xue}

\author[inst2]{Emmanuelle A. Marquis}

\affiliation[inst2]{organization={Dept.~of Materials Science and Engineering, University of Michigan},
            city={Ann Arbor},
            state={MI},
            country={USA}}
            
\author[inst1]{{Wolfgang Windl}\texorpdfstring{\corref{cor1}}}
\cortext[cor1]{Corresponding author}

\begin{abstract}
In atom probe tomography (APT), atoms from the surface of a needle shape specimen are evaporated under a high electric field and analyzed via time of flight mass spectrometry and position sensitive detection. 3D reconstruction of the atom positions follows a simple projection law, which can sometimes lead to artifacts due to deviation from an assumed ideal evaporation sequence. Here, we revisit the evaporation behavior of [001]-oriented $\gamma$-TiAl using a full-dynamics simulation approach empowered by molecular dynamics. Without any knowledge of charge states or assumptions about evaporation fields, we successfully reproduced the lack of distinct Al and Ti layers observed in reconstructions of experimental data which is traditionally attributed to the retention of Al on the evaporating surface. We further showed that a step-wise bond breaking process of Ti in contrast to the simultaneous bond breaking of Al explains the seemingly counterintuitive preferential evaporation of the strongly bonded Ti atoms.
\end{abstract}


\end{frontmatter}


\section{Background}
\label{sec:TiAl-background}

Atom probe tomography (APT) is known as a very unique characterization technique that possesses the ability to resolve both atomic structure and chemical composition of materials in three dimensions. In APT, surface atoms of a sharp needle-shape specimen are field-evaporated under a high electric field, which is created by a high voltage applied to the specimen. A two-dimensional detector records the sequence and positions of the arriving atoms. Along with the chemical identity information recovered by mass spectrometry, a 3D distribution of atoms in the evaporated volume can be generated based on a simple projection law \cite{bas1995ageneral}.

The standard reconstruction protocol assumes that the evaporating specimen always keeps a perfect hemispherical shape, which requires atoms to leave the surface in an ideal sequence. However, the evaporation of materials with layered structures often shows different rates of evaporation of the different layers with one tending to stick to the surface while the other evaporates readily \cite{marquis2011evolution}. This sort of behavior is typically explained by a hand-waving argument: different atoms have different evaporation fields, which disturbs the ideal evaporation sequence and results in positioning errors during the reconstruction of certain atomic planes.

$\gamma$-TiAl is an excellent system for examining such errors in the observed layer sequences in APT. Experimentally it is found that the subsequent, distinct Al and Ti planes, as one would expect from the crystallographic structure of TiAl (Fig.~\ref{fig:rec}(a)) are merged by the reconstruction process leading to half the number of planes, each containing Al and Ti (Fig.~\ref{fig:rec}(c)). 

Field ion microscopy observations \cite{wesemann1995apfim, kim1997ap-fim, boll2007investigation} typically revealed that the Al planes constitute the top layer, while the following Ti plane is only slightly larger than the Al top layer but much smaller than the following Al layer as shown in Fig.~\ref{fig:surface}(a). These observations have largely been interpreted as the result of Al being retained on the surface while Ti readily evaporates once uncovered, due to the lower evaporation of Ti as compared to that of Al. However, Lefebvre et al.\ came to the opposite conclusion proposing that Ti has a higher evaporation field than Al based on based on APT reconstructed data where Ti atoms were positioned slightly higher than Al atoms within each mixed plane \cite{lefevre2002field}. Subsequently, Boll et al.\ performed simulations in order to reproduce their own experimental observation of Ti evaporating more easily than Al \cite{boll2013interpretation}. They used M\"uller's formula for the evaporation field, from which they stated to have found a higher evaporation field for Al when assuming for both Al and Ti a +2 charge state as observed experimentally. However, their calculated evaporation fields are not reported in the paper, and it is not possible to reproduce them since the bond energy $\Lambda$ of Ti-Al used to calculate the sublimation energy is $\Lambda_{\rm Al-Ti} = 0.70$~eV. Compared to the energies of the homoelemental bonds, $\Lambda_{\rm Al-Al} = 0.33$~eV and $\Lambda_{\rm Ti-Ti} = 1.07$~eV, the value of 0.70~eV results in an ordering energy of zero, which suggests Ti and Al form an ideal solid solution rather than an ordered intermetallic. Our own DFT-values for these bond energies, which will be discussed later, are in stark contrast to either value, with $\Lambda_{\rm Al-Ti} = 0.86$~eV, $\Lambda_{\rm Al-Al} = 0.58$~eV, and $\Lambda_{\rm Ti-Ti} = 0.91$~eV and an ordering energy of $-0.12$~eV in the common range of intermetallics.

In order to sort through these contradictory findings, we revisit here the field evaporation sequence of $[001]$-oriented TiAl by applying the full-dynamics field evaporation simulation approach TAPSim-MD \cite{qi2022abinitio}. Our approach evaporates atoms in the specimen as a result of competition between interatomic forces and electrostatic forces, and gives an ``\textit{ab-initio}'' prediction of evaporation events as well as atom trajectories. Here for the example of $[001]$-oriented TiAl, we show that without any knowledge of charge states, TAPSim-MD successfully reproduces surface morphologies observed by FIM and the mixed-layer reconstructed structure. Our results suggest that Al indeed possesses a higher evaporation field than Ti, which is in agreement with most previous conclusions from experiments. However, in contrast to previous suggestions, we find from analyzing the atomic trajectories that Ti is preferentially evaporated due to its two-step bond breaking process, which is easier to happen in comparison to the evaporation process for Al, where all bonds are simultaneously broken.

\section{Methods}\label{sec2}

Simulations were conducted using the full-dyanmics APT simulation tool TAPSim-MD \cite{qi2022abinitio}, which combines the traditional finite element APT simulation package TAPSim \cite{oberdorfer2013afullscale, tapsim} with the classical molecular dynamics approach as implemented in LAMMPS \cite{plimption1995fast, lammps}. A [001] oriented virtual tip is built in cylinder shape with 8 nm in radius and 24 nm in height. 
The virtual voltage in the simulation is first increased by 10 V after a set of MD runs until 420 V, and then increased by 2 V after a set of MD runs until an evaporation event is triggered. The voltage never decreases to ensure a computationally efficient data collection process. The employed interatomic potential is an embedded atom model potential from Zope and Mishin \cite{zope2003interatomic}.

For comparison, we also performed traditional TAPSim simulations for the same tip structure. Since no MD is included in this method, an evaporation criterion is used to determine the evaporation event which is based on the magnitude of field-induced forces and empirical evaporation fields. Since the hypothesis to test is if Al or Ti should have the lower evaporation field, we use type 1 and type 2 to tag the atoms in the two layers. We set the evaporation field of type 1 to 2.6 V/{\AA} as the high field, and that of type 2 to 1.9 V/{\AA} as the low field. Details about the TAPSim approach and applications can be found in Refs.~\cite{oberdorfer2013afullscale, oberdorfer2015applications}.

Specimens for APT were prepared using a standard lift-out and sharpening process performed with a Thermo Fisher Helios 650 Nanolab scanning electron microscope/focused ion beam (SEM/FIB) from a Ti-52Al-2Nb (at.\%) ingot in as-cast condition. A $\langle 001\rangle$-oriented grain was identified by electron backscatter diffraction (EBSD) using a Tescan RISE SEM. Data post-processing used the Orientation Imaging Microscopy (OIM) software version 9 to address the pseudosymmetry issue through spherical indexing and real-space refinement. The orientation of the $\langle 001\rangle$ grain was further verified by selective area diffraction (SAD) with a Thermo Fisher Talos F200X G2 scanning/transmitting electron microscope (S/TEM) using the c/a ratio of tetragonal $\gamma$-TiAl (L1$_0$) and the positions of the superlattice spots. APT experiments were conducted using a Cameca LEAP 5000XR instrument. Data collection was conducted with a base sample temperature of 50 K and a detection rate of 1\% in laser mode. A laser energy of 50 pJ and a pulse frequency of 200 kHz were used. The APT data was reconstructed using the Cameca AP Suite software 6.3. Values of the image compression factor were estimated using the desorption map indexing by APS-integrated OIM and were found to be 1.58-1.59. The field factor was optimized between 5.95 and 9.40 to create a (001) plane spacing close to 0.4 nm comparable to that of $\gamma$-TiAl. The measured compositions were Ti-(53.4-54.0 at.\%)Al-1.7 at.\%Nb, close to the nominal composition.

\section{Results and Discussion}
\subsection{Reconstruction and Surface Morphology}
In this section, we undertake a comparative analysis by juxtaposing various simulation outcomes with experimental observations, focusing on the standard reconstruction outcome and the surface morphology of the tip.

\begin{figure}[!htb]
    \includegraphics[width=0.95\linewidth]{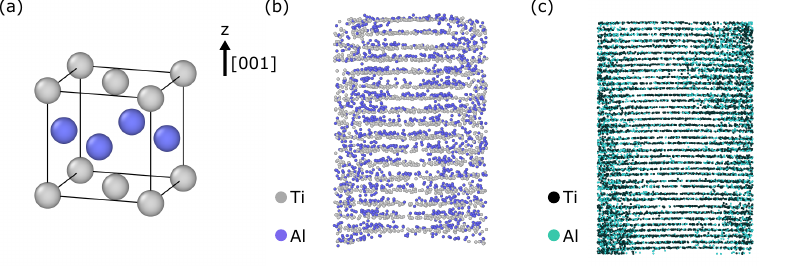}
    \centering
    \caption{(a) is the structure of $\gamma$-TiAl intermetallic compound. (b) is the standard reconstruction result of simulation data from TAPSim-MD. (c) is the experimental reconstruction result of TiAl.}
    \label{fig:rec}
\end{figure}

As mentioned before, the reconstructed structure of $[001]$ oriented TiAl has mixed Ti/Al planes instead of alternate Ti and Al planes in the crystal as shown in Fig.~\ref{fig:rec}(a). Our simulated reconstruction result (Fig.\ref{fig:rec}(b)) is consistent with the experimental observation shown in Fig.~\ref{fig:rec}(c). In both cases, there is no discernible alternately layered structure. 

\begin{figure}[!htb]
    \includegraphics[width=0.95\linewidth]{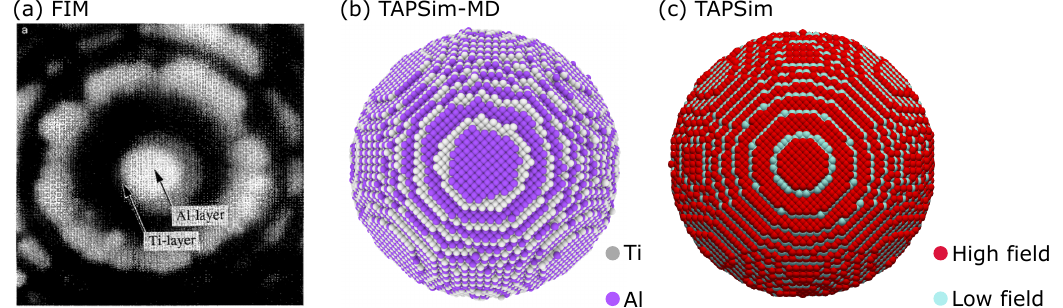}
    \centering
    \caption{(a) Field-ion image of the (001) pole of a Ti$_{51}$Al$_{46}$Cr$_{3}$ sample, reprinted from Ref.~\cite{wesemann1995apfim}. The brighter (dimmer) contrast is Al (Ti). (b) Top view of a TiAl tip simulated by TAPSim-MD. Ti (Al) atoms are grey (violet). (c) Top view of the tip from (b), simulated by traditional TAPSim. Atoms with higher (lower) evaporation field are red (cyan).}
    \label{fig:surface}
\end{figure}

As illustrated in the field-ion image in Fig.\ref{fig:surface}(a), the size of the Ti layer closely matches that of the Al layer situated above it, whereas it is significantly smaller than the Al layer beneath. Our molecular dynamics (MD) simulation, shown in Fig.\ref{fig:surface}(b), indeed replicates this coupled plane evaporation pattern. Remarkably, a traditional TAPSim simulation, without the inclusion of MD, also yields the same pattern, as shown in Fig.~\ref{fig:surface}(c), where the low-field cyan plane has a similar size to the high-field red plane on top of it.

The comparison between the simulation outcomes of TAPSim-MD and traditional TAPSim reveals that Al atoms are likely to possess a higher evaporation field, while Ti atoms are expected to exhibit a lower one. Consequently, Al atoms tend to remain on the specimen, whereas Ti atoms have a greater propensity for evaporation. This result is in agreement with most previous analyses in the literature \cite{cerezo1996materials, wesemann1995apfim, kim1997ap-fim, boll2007investigation}. 

\subsection{Distinct Atom Trajectories and Bond-breaking Processes}
As demonstrated previously, TAPSim-MD's prediction of the bilayer surface morphology along with the evaporation sequence for TiAl agree well with experimental observations without any \textit{ad hoc} assumptions about evaporation fields or prior knowledge of charge states. The full dynamics of TAPSim-MD allows us now to study the underlying processes of the evaporation sequence from a dynamic perspective by analyzing forces and atom trajectories. 

\begin{figure}[!htb]
    \includegraphics[width=0.95\linewidth]{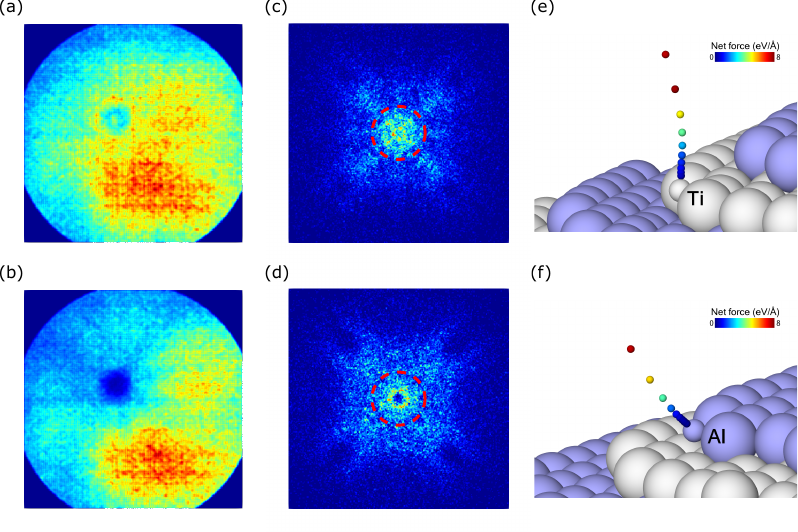}
    \centering
    \caption{Field evaporation maps from the experiment and TAPSim-MD simulation for Ti atoms ((a) and (c)) and Al atoms ((b) and (d)). The warmer the color, the higher the intensity. Red circles in (c) and (d) depict a similar field of view compared to the experimental results in (a) and (b). Further shown are examples of (e) Ti and (f) Al atom trajectories during evaporation, respectively. Ti (Al) atoms are shown in grey (violet). The equilibrium positions of the evaporating atoms are indicated by smaller balls with the same color scheme. The atom trajectories are described by a series of small dots colored by the magnitude of net forces at different instances. The warmer the color, the larger the net force.}
    \label{fig:traj}
\end{figure}

Fig.~\ref{fig:traj}(a)-(d) depict markedly different field evaporation maps for Ti and Al: the central pole in the Ti map is more densely populated, while it is depleted in the Al map, which can be observed in both experimental and simulated results. Furthermore, in the simulated evaporation maps, the zone lines in the Ti map (c) exhibit heightened intensity, whereas a slight depletion is observed in the Al map (d). These pronounced dissimilarities in the evaporation maps strongly suggest that Ti and Al follow distinctly different launch trajectories.
Indeed, our observed launch trajectories for Ti and Al, as shown in Fig.~\ref{fig:traj}(c) and (d), confirm this divergence: the evaporating Ti atom initially ascends vertically along the $z$-direction, while the evaporating Al atom departs at an angle of approximately 45$^\circ$ between the $z$-axis and the $xy$-plane.

Considering that the electrostatic forces should not depend on the chemical identities of evaporating atoms for equal charge states but on the geometry, the distinct atom launch trajectories of Ti and Al indicate that the desorption or bond breaking processes of the two species are distinctly different. 
Based solely on bond strength, Al would be expected to have weaker bonds than Ti. Nevertheless, it exhibits a higher evaporation field than Ti, despite Ti possessing relatively strong bonds. As mentioned, we find the bond strengths of $-0.58$ eV, $-0.86$ eV, and $-0.91$ eV for Al-Al, Ti-Al and Ti-Ti bonds, respectively, from DFT calculations . 
While it is well-established that a comprehensive analysis of real evaporation processes should consider multiple factors, including total ionization energy and work function, our MD simulation employs induced charges to calculate electrostatic forces and otherwise solely depends on interatomic interactions. Consequently, the preferential evaporation of strongly bonded Ti atoms can be attributed to the manner in which these bonds are broken during the evaporation process.

\begin{figure}[!htb]
    \includegraphics[width=0.9\linewidth]{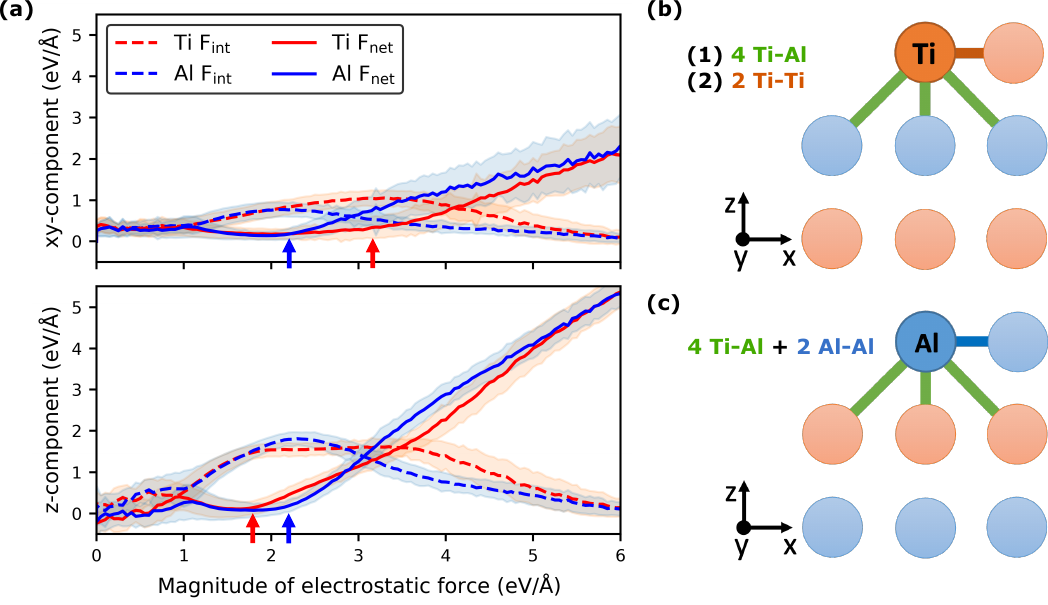}
    \centering
    \caption{(a) $xy$- and z-components of interatomic forces (dashed lines) and net forces (solid lines) vs.\ the magnitude of the electrostatic force during field evaporation in TAPSim-MD simulation for an 8 nm-radius TiAl tip. Each data point is an average over 3098 Ti atoms or 2871 Al atoms. The standard deviations are indicated by shadows. The critical points are indicated by blue or red arrows for Ti or Al. The schematic bonding condition and bond breaking sequence of evaporating atoms are shown for a Ti atom (orange) in (b) and for an Al atom (blue) in (c).}
    \label{fig:bond}
\end{figure}

In order to understand the bond breaking process, we first examine the $xy$- and $z$-components of the interatomic force and net force as functions of the electrostatic force induced by the applied voltage. As shown in fig.~\ref{fig:bond}(a), when the electrostatic force induced on an atom is rather small (approximately less than 1 eV/\AA), it is balanced by the interatomic force, resulting in an almost zero net force in both the $xy$-plane and the $z$-direction, which makes the atom vibrate around its equilibrium position. However, as the electrostatic force increases, it begins to overcome the interatomic force and consequently gives rise to a continuously growing instead of oscillating net force, where the atom starts to evaporate. With regard to Al (blue lines), the continuous growth of the net force (solid lines) in both $xy$-plane (upper panel) and $z$-direction (lower panel) start when the electrostatic force is about 2.2 eV/\AA, as indicated by blue arrows. As for Ti atoms (red lines), the continuous growth of the lateral net force starts around an electrostatic force of 3.2 eV/{\AA}, while the growth of the $z$-component starts around an electrostatic force of 1.8 eV/\AA, as indicated by red arrows. The simultaneous growth of the lateral and $z$-components of the net force of Al suggests that bond breaking in the $xy$-plane (2 Al-Al bonds) and $z$-direction (4 Ti-Al bonds) should occur around the same time. In contrast, for Ti, the in-plane and the out-of-plane bond breaking start at distinctly different times, first in the $z$-direction (breaking 4 Ti-Al bonds at an electrostatic force of $\sim$1.8 eV/\AA), and second in the lateral direction (breaking 2 Ti-Ti bonds at an electrostatic force of $\sim$3.2 eV/\AA). Schematic plots of the bond breaking processes for the two species along these findings are shown in Fig.~\ref{fig:bond}(b)-(c).

\begin{figure}[!htb]
    \includegraphics[width=\linewidth]{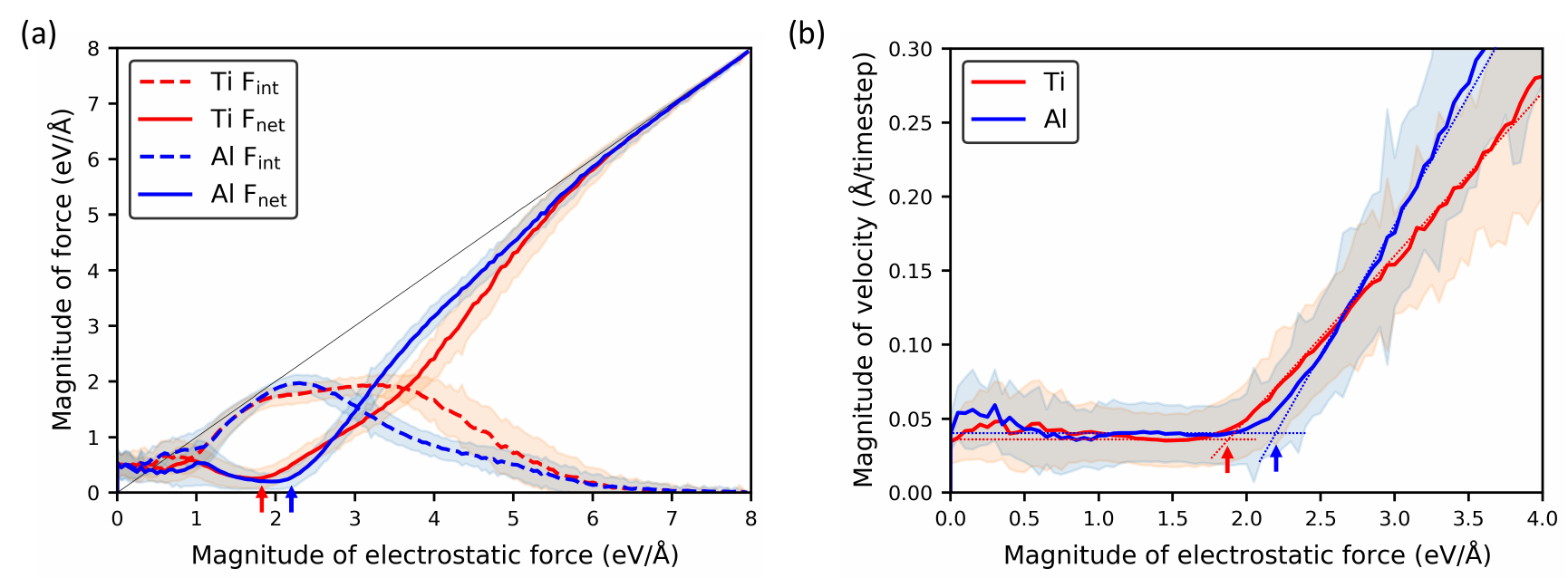}
    \centering
    \caption{(a) Magnitude of interatomic and net force vs magnitude of electrostatic force. The thin black solid line indicates y=x. (b) Magnitude of velocity vs magnitude of electrostatic force. Each data point is averaged over 3098 Ti atoms and 2871 Al atoms. The standard deviations are indicated by shadows. The critical points are indicated by blue or red arrows for Ti or Al.}
    \label{fig:f_v}
\end{figure}

In order to show why the two-step bond-breaking process of Ti results in its preferential evaporation, we examine the magnitude of both the total net force and the resulting velocity as functions of the electrostatic force.  Figure~\ref{fig:f_v}(a) clearly shows that the net force acting on the Ti atom (red line) doesn't exhibit continuous growth until the electrostatic force surpasses approximately 1.8 eV/{\AA}, indicated by a red arrow, while this point is reached later for Al (blue line), at around 2.2 eV/{\AA}, indicated by a blue arrow. 
Consequently, when subjected to the same magnitude of electrostatic forces, it becomes easier to initiate the evaporation of a Ti atom. Along with that, the two-step bond-breaking process of Ti is also reflected on the growth of the interatomic force (red dashed line), where two local maxima can be found at around 1.9 eV/{\AA} and 3.3 eV/{\AA} electrostatic force, corresponding to the first and second bond-breaking events of Ti respectively. In contrast, there is only one interatomic force maximum of Al (blue dashed line) at around 2.3 eV/{\AA} electrostatic force, which is associated with the one-step bond-breaking process of Al. As a side note, the electrostatic force and internal force can not completely offset each other after the atom leaves its equilibrium position as a result of the different anisotropic features of the interatomic force and electrostatic force \cite{qi2023origin}, which explains why the growth of the interatomic force deviates from the line of $y=x$ at the initial stage of evaporation.
 
The conclusion of preferential evaporation of Ti can also be drawn from analyzing the velocities shown in Fig.~\ref{fig:f_v}(b). For both Al and Ti, the atoms exhibit vibrational motion around their equilibrium positions with nearly zero velocity before the electrostatic force starts to dominate over the interatomic force. As the electrostatic force intensifies, a critical point is reached at approximately 1.8 eV/{\AA} for Ti and 2.2 eV/{\AA} for Al, quantified by the electrostatic force at the intersection of the linear-acceleration range with the constant, low-velocity range. At this point, the interatomic force cannot balance the electrostatic force anymore, and the velocity starts to linearly grow with the increasing electrostatic force. Once again, when considering the application of an equivalent magnitude of electrostatic forces, we find that the evaporation of a Ti atom is easier than that of Al, equivalent to a lower evaporation field of Ti compared to Al. Interestingly, the velocity analysis and especially the determined intersection point suggests that it is only the first bond-breaking event in Ti that seems to correlate with the effective evaporation field rather than the second one.

\section{Conclusion}\label{sec4}

In this paper, we have revisited the disrupted evaporation sequence in [001]-oriented $\gamma$-TiAl, a phenomenon that often leads to erroneous results in reconstruction due to the different evaporation fields exhibited by different species. 
Employing our full-dynamics approach, we provide an \textit{ab initio} confirmation that Ti requires a lower electrostatic force along with a lower evaporation field to evaporate, a finding that aligns well with the majority of experimental observations documented in the literature. More importantly, through a comprehensive examination of atom trajectories and forces, we have unveiled a step-wise bond-breaking process for Ti, in contrast to the simultaneous bond-breaking process observed for Al. This discovery offers a compelling explanation for the seemingly counterintuitive preferential evaporation of the strongly bonded Ti atoms.

\section{Acknowledgement}

This project has been sponsored by AFOSR (PM Dr. Ali Sayir) under Award No. FA9550-14-1-0249 and FA9550-19-1-0378. Computing resources were supplied and maintained by the Ohio Supercomputer Center \cite{OhioSupercomputerCenter1987} under grant number PAA0010. Results of the simulations were visualized with the programs OVITO \cite{ovito}, ParaView \cite{paraview}, and Python library Matplotlib \cite{matplotlib}.


\begin{thebibliography}{10}
\expandafter\ifx\csname url\endcsname\relax
  \def\url#1{\texttt{#1}}\fi
\expandafter\ifx\csname urlprefix\endcsname\relax\def\urlprefix{URL }\fi
\expandafter\ifx\csname href\endcsname\relax
  \def\href#1#2{#2} \def\path#1{#1}\fi

\bibitem{bas1995ageneral}
P.~Bas, A.~Bostel, B.~Deconihout, D.~Blavette, A general protocol for the reconstruction of 3d atom probe data, Applied Surface Science 87-88 (1995) 298--304.
\newblock \href {https://doi.org/https://doi.org/10.1016/0169-4332(94)00561-3} {\path{doi:https://doi.org/10.1016/0169-4332(94)00561-3}}.

\bibitem{marquis2011evolution}
E.~A. Marquis, B.~P. Geiser, T.~J. Prosa, D.~J. Larson, Evolution of tip shape during field evaporation of complex multilayer structures, Journal of Microscopy 241~(3) (2011) 225--233.
\newblock \href {https://doi.org/https://doi.org/10.1111/j.1365-2818.2010.03421.x} {\path{doi:https://doi.org/10.1111/j.1365-2818.2010.03421.x}}.

\bibitem{wesemann1995apfim}
J.~Wesemann, G.~Frommeyer, M.~Kreuss, \href{https://www.sciencedirect.com/science/article/pii/0169433294004838}{Apfim investigations on ordered $\gamma$-tial using single-layer detection method}, Applied Surface Science 87-88 (1995) 179--184, proceedings of the 41st International Field Emission Symposium.
\newblock \href {https://doi.org/https://doi.org/10.1016/0169-4332(94)00483-8} {\path{doi:https://doi.org/10.1016/0169-4332(94)00483-8}}.
\newline\urlprefix\url{https://www.sciencedirect.com/science/article/pii/0169433294004838}

\bibitem{kim1997ap-fim}
S.~Kim, G.~Smith, \href{https://www.sciencedirect.com/science/article/pii/S0921509397005868}{Ap-fim investigation on $\gamma$-based titanium aluminides}, Materials Science and Engineering: A 239-240 (1997) 229--234, 4th Conference on High-Temperature Intermetallics.
\newblock \href {https://doi.org/https://doi.org/10.1016/S0921-5093(97)00586-8} {\path{doi:https://doi.org/10.1016/S0921-5093(97)00586-8}}.
\newline\urlprefix\url{https://www.sciencedirect.com/science/article/pii/S0921509397005868}

\bibitem{boll2007investigation}
T.~Boll, T.~Al-Kassab, Y.~Yuan, Z.~Liu, \href{https://www.sciencedirect.com/science/article/pii/S0304399107000605}{Investigation of the site occupation of atoms in pure and doped tial/ti3al intermetallic}, Ultramicroscopy 107~(9) (2007) 796--801.
\newblock \href {https://doi.org/https://doi.org/10.1016/j.ultramic.2007.02.011} {\path{doi:https://doi.org/10.1016/j.ultramic.2007.02.011}}.
\newline\urlprefix\url{https://www.sciencedirect.com/science/article/pii/S0304399107000605}

\bibitem{lefevre2002field}
W.~Lefebvre, A.~Loiseau, A.~Menand, \href{https://www.sciencedirect.com/science/article/pii/S0304399102000700}{Field evaporation behaviour in the $\gamma$ phase in ti–al during analysis in the tomographic atom probe}, Ultramicroscopy 92~(2) (2002) 77--87.
\newblock \href {https://doi.org/https://doi.org/10.1016/S0304-3991(02)00070-0} {\path{doi:https://doi.org/10.1016/S0304-3991(02)00070-0}}.
\newline\urlprefix\url{https://www.sciencedirect.com/science/article/pii/S0304399102000700}

\bibitem{boll2013interpretation}
T.~Boll, T.~Al-Kassab, Interpretation of atom probe tomography data for the intermetallic {T}i{A}l+{N}b by means of field evaporation simulation, Ultramicroscopy 124 (2013) 1--5.
\newblock \href {https://doi.org/https://doi.org/10.1016/j.ultramic.2012.09.003} {\path{doi:https://doi.org/10.1016/j.ultramic.2012.09.003}}.

\bibitem{qi2022abinitio}
J.~Qi, C.~Oberdorfer, W.~Windl, E.~A. Marquis, \href{https://link.aps.org/doi/10.1103/PhysRevMaterials.6.093602}{Ab initio simulation of field evaporation}, Phys. Rev. Materials 6 (2022) 093602.
\newblock \href {https://doi.org/10.1103/PhysRevMaterials.6.093602} {\path{doi:10.1103/PhysRevMaterials.6.093602}}.
\newline\urlprefix\url{https://link.aps.org/doi/10.1103/PhysRevMaterials.6.093602}

\bibitem{oberdorfer2013afullscale}
C.~Oberdorfer, S.~M. Eich, G.~Schmitz, A full-scale simulation approach for atom probe tomography, Ultramicroscopy 128 (2013) 55--67.
\newblock \href {https://doi.org/10.1016/j.ultramic.2013.01.005} {\path{doi:10.1016/j.ultramic.2013.01.005}}.

\bibitem{tapsim}
{APT} {S}oftware {TAPSim}, \url{https://www.imw.uni-stuttgart.de/mp/forschung/atom_probe_RD_center/software/}, accessed: 2021-09-19.

\bibitem{plimption1995fast}
S.~Plimpton, Fast parallel algorithms for short-range molecular dynamics, Journal of Computational Physics 117~(1) (1995) 1--19.
\newblock \href {https://doi.org/https://doi.org/10.1006/jcph.1995.1039} {\path{doi:https://doi.org/10.1006/jcph.1995.1039}}.

\bibitem{lammps}
{LAMMPS} {M}olecular {D}ynamics {S}imulator, \url{http://lammps.org}, accessed: 2021-09-19.
\newblock \href {https://doi.org/10.5281/zenodo.3726416} {\path{doi:10.5281/zenodo.3726416}}.

\bibitem{zope2003interatomic}
R.~R. Zope, Y.~Mishin, \href{https://link.aps.org/doi/10.1103/PhysRevB.68.024102}{Interatomic potentials for atomistic simulations of the ti-al system}, Phys. Rev. B 68 (2003) 024102.
\newblock \href {https://doi.org/10.1103/PhysRevB.68.024102} {\path{doi:10.1103/PhysRevB.68.024102}}.
\newline\urlprefix\url{https://link.aps.org/doi/10.1103/PhysRevB.68.024102}

\bibitem{oberdorfer2015applications}
C.~Oberdorfer, S.~M. Eich, M.~Lütkemeyer, G.~Schmitz, Applications of a versatile modelling approach to 3d atom probe simulations, Ultramicroscopy 159 (2015) 184--194.
\newblock \href {https://doi.org/https://doi.org/10.1016/j.ultramic.2015.02.008} {\path{doi:https://doi.org/10.1016/j.ultramic.2015.02.008}}.

\bibitem{cerezo1996materials}
A.~Cerezo, D.~Gibuoin, S.~Kim, S.~Sijbrandij, F.~Venker, P.~Warren, J.~Wilde, G.~Smith, Materials applications of an advanced 3-dimensional atom probe, JOURNAL DE PHYSIQUE IV 6~(C5) (1996) 205--210, 43rd International Field Emission Symposium (IFES96), MOSCOW, RUSSIA, JUL 14-19, 1996.
\newblock \href {https://doi.org/10.1051/jp4:1996533} {\path{doi:10.1051/jp4:1996533}}.

\bibitem{qi2023origin}
J.~Qi, C.~Oberdorfer, E.~A. Marquis, W.~Windl, \href{https://www.sciencedirect.com/science/article/pii/S1359646223001306}{Origin of enhanced zone lines in field evaporation maps}, Scripta Materialia 230 (2023) 115406.
\newblock \href {https://doi.org/https://doi.org/10.1016/j.scriptamat.2023.115406} {\path{doi:https://doi.org/10.1016/j.scriptamat.2023.115406}}.
\newline\urlprefix\url{https://www.sciencedirect.com/science/article/pii/S1359646223001306}

\bibitem{OhioSupercomputerCenter1987}
O.~S. Center, \href{http://osc.edu/ark:/19495/f5s1ph73}{Ohio supercomputer center} (1987).
\newline\urlprefix\url{http://osc.edu/ark:/19495/f5s1ph73}

\bibitem{ovito}
A.~Stukowski, \href{https://doi.org/10.1088/0965-0393/18/1/015012}{Visualization and analysis of atomistic simulation data with {OVITO}{\textendash}the open visualization tool}, Modelling and Simulation in Materials Science and Engineering 18~(1) (2009) 015012.
\newblock \href {https://doi.org/10.1088/0965-0393/18/1/015012} {\path{doi:10.1088/0965-0393/18/1/015012}}.
\newline\urlprefix\url{https://doi.org/10.1088/0965-0393/18/1/015012}

\bibitem{paraview}
U.~Ayachit, The ParaView Guide: A Parallel Visualization Application, Kitware, Inc., Clifton Park, NY, USA, 2015.

\bibitem{matplotlib}
J.~D. Hunter, Matplotlib: A 2d graphics environment, Computing in Science \& Engineering 9~(3) (2007) 90--95.
\newblock \href {https://doi.org/10.1109/MCSE.2007.55} {\path{doi:10.1109/MCSE.2007.55}}.

\end{thebibliography}

\end{document}